\documentclass[aps,showpacs,eqsecnum,superscriptaddress,preprintnumbers]{revtex4}

\usepackage{amsmath}
\usepackage{amssymb}
\usepackage{latexsym}
\usepackage{graphicx}
\usepackage{times}

\newcommand{\scri}{\cal I}
\newcommand{\LIE}[1]{{\cal L}_{#1}}

\begin{document}
\preprint{AEI-2005-004}

\title{Gravitational wave extraction based on Cauchy-characteristic
extraction and characteristic evolution}

\author{Maria Babiuc}
\affiliation{Department of Physics and Astronomy, University of
  Pittsburgh, Pittsburgh, PA 15260, US}

\author{B\'{e}la Szil\'{a}gyi}
\affiliation{Department of Physics and Astronomy, University of
  Pittsburgh, Pittsburgh, PA 15260, US}
\affiliation{Max-Planck-Institut f\"ur Gravitationsphysik,
  Albert-Einstein-Institut, Am M\"uhlenberg 1, D-14476 Golm, Germany}

\author{Ian Hawke}
\affiliation{Max-Planck-Institut f\"ur Gravitationsphysik,
  Albert-Einstein-Institut, Am M\"uhlenberg 1, D-14476 Golm, Germany}
\affiliation{School of Mathematics, University of Southampton,
  Southampton SO17 1BJ, UK}

\author{Yosef Zlochower} 
\affiliation{Department of Physics and Astronomy, and Center for
  Gravitational Wave Astronomy, The University of Texas at
  Brownsville, Brownsville, TX 78520, US}

\date{$ $Date: 2005/09/20 14:30:02 $ $}


\begin{abstract}
  We implement a code to find the gravitational news at future null
  infinity by using data from a Cauchy code as boundary data for a
  characteristic code. This technique of {\it Cauchy-characteristic
    Extraction} (CCE) allows for the unambiguous extraction of
  gravitational waves from numerical simulations. We first test the
  technique on non-radiative spacetimes: Minkowski spacetime, 
  perturbations of Minkowski spacetime and static black hole spacetimes 
  in various gauges. We show the convergence and limitations of the 
  algorithm and illustrate its success in cases where other wave 
  extraction methods fail.
  We further apply our techniques to a standard radiative test case for wave 
  extraction: a linearized Teukolsky wave, presenting our results 
  in comparison to the Zerilli technique and we argue for the advantages of 
  our method of extraction. 
\end{abstract}

\maketitle

\section{Introduction}
\label{sec:intro}

The importance of gravitational waveform templates for gravitational wave
detectors implies a need for accurate 3D numerical simulations of
isolated sources such as binary black hole mergers.  These simulations
are often done with Cauchy codes based on a ``3+1'' slicing of
spacetime. With the slicing conditions most commonly used in numerical
simulations, two problems arise: artificial boundary conditions must
be placed on the computational domain, and information such as the
gravitational news cannot be extracted at future null infinity
$\scri^+$. 

To avoid these problems two possible approaches have been suggested.
One way is to evolve the entire spacetime. For example, a {\em hyperboloidal} 
slicing of the spacetime~\cite{Huebner01, Frauendiener04, Husa01} allows information to propagate to $\scri^+$ in finite time. Conformal compactifications,
such as those suggested by the conformal field equations~\cite{Friedrich81a,
Friedrich81b} or employed in~\cite{Fodor04} allow $\scri^+$ to be located
at a finite computational coordinate in a regular way.
Another example is characteristic evolution based on a Bondi-Sachs line element
, which gives a natural description of the radiation-region of spacetime
extending to $\scri^+$. Characteristic numerical codes have been used to study tail decay~\cite{Gomez94b}, critical phenomena~\cite{Stewart96,Garfinkle95,
Husa2000b,Husain01,Husain03}, singularity structure~\cite{Gnedin93,Brady95,
Burko97,Hod98} and fluid collapse~\cite{Bishop99,Siebel02,Siebel03}, to list 
just a few examples.

Unfortunately, characteristic methods suffer from the appearance of
caustics in the inner strong field region. The problem of caustics can be
avoided by evolving the strong field region with a standard Cauchy
slicing whilst using the characteristic approach for the exterior.
This technique of {\em Cauchy-characteristic matching} (CCE) has
proved successful in numerical evolution of the spherically symmetric
Klein-Gordon-Einstein field equations~\cite{Gomez97b}, for 3D
non-linear wave equations~\cite{Bishop97c} and for the linearized
harmonic Einstein system~\cite{Szilagyi02b}.

The second way of avoiding problems with standard Cauchy codes is a 
perturbative approach. The standard way of extracting gravitational waves from
a numerical simulation is based on perturbation theory, either using
the quadrupole formula or first order gauge invariant formalisms based on 
the work of Zerilli and Moncrief~\cite{Zerilli70,Moncrief74}. 
The vast majority of waveform templates are currently based upon these 
approaches. 
Numerical codes able to solve a generalization of the Zerilli equation to a 
three dimensional Cartesian coordinate system and to extract the gravitational 
signal are reported in~\cite{Rezzolla97a, Shibata97e, Nakao01, Yo01}.
Extraction of gravitational waves based on the Zerilli-Moncrief formalism in 
fully three-dimensional simulations are presented also in~\cite{Baiotti04b, 
Sperhake2005a}.

 Perturbative methods have been also used to provide boundary conditions at the
outer boundary of the Cauchy grid. This approach of {\em Cauchy-perturbative
  matching} has been implemented numerically in~\cite{Rupright98,Rezzolla99a,
Abrahams97a}. All those works are impressive, but discrepancies between the
results of a perturbative approach and the full non-linear theory
cannot be determined without solving the fully non-linear problem, although
error estimates have been made~\cite{Bishop98a}.

Of these approaches, CCM has many appealing properties. 
The characteristic description of the exterior,
particularly at $\scri^+$, allows for a natural extraction of
gravitational information such as the news. 
Matching to a standard Cauchy code in the interior implies that the methods 
employed in current 3D numerical simulations may immediately be used. As both
Cauchy and characteristic approaches have been well tested and commonly used, 
CCM is a natural way of avoiding the potential problems of caustics in the 
characteristic region and artificial boundaries in the Cauchy region.

In this work we take a step towards the full CCM method by using a Cauchy code 
to provide boundary data for a characteristic code which propagates the 
solution to $\scri^+$ to extract the waveform. This procedure of 
{\em Cauchy-characteristic extraction} (CCE) allows the computation 
of gravitational waves in an unambiguous fashion. 
CCE was succesfully implemented in the quasispherical approximation 
in~\cite{Bishop96}; here we demonstrate it in the fully non-linear case. 

In this work the outer region extending to $\scri^+$ is numerically evolved 
by the Pitt Null Code~\cite{Lehner99a, Lehner98, Bishop96, Gomez97b, Gomez97a, 
Bishop97b, Gomez01,Winicour98, Winicour05}. The link between the Cauchy and the
characteristic modules is done by a non-linear 3D CCE algorithm~\cite{Bishop96,
Bishop98b, Winicour98, Winicour05, Szilagyi00}. At the outer edge of the 
characteristic grid the Bondi news is computed~\cite{Bishop97b, Lehner98, 
Zlochower02, Zlochower03}. 
We have imported the CCE code into the Cactus computational infrastructure~\cite{Allen01b, Allen01c, Cactusweb, Goodale02a}. 
Within this infrastructure we have been able to use two separate Cauchy codes 
implementing the BSSN and harmonic $3+1$ formulations of the Einstein 
equations. This allows us to test the robustness of the CCE approach.  

We compare CCE with the Zerilli extraction technique~\cite{Zerilli70, 
Moncrief74, Camarda97c} for both non-radiative and radiative spacetimes.  
The Zerilli formalism is based on the paper of Nagar and Rezzolla~\cite{Nagar05} 
which reviews and collects the relevant expressions related to the 
Regge-Wheeler and Zerilli equations for the odd and even-parity perturbations 
of a Schwarzschild spacetime. The conventions presented in their review are 
implemented in the Wave Extract code, included in the Cactus computational 
toolkit open source infrastructure~\cite{Cactusweb}.  

For non-radiative spacetimes, the comparison gives results 
consistent with reference~\cite{Bishop98a}: the CCE algorithm is 
${\cal O}(\Delta^2)$ accurate (where $\Delta$ is the computational grid-step), 
while the accuracy of the perturbative (Zerilli) approach is ${\cal O}(r^{-2})$
 where $r$ is the radius of the wave extraction sphere. 
Since the signal is ${\cal O}(r^{-1})$, this implies an ${\cal O}(r^{-1})$ 
relative error in the Zerilli approach.
As a result, it is shown in~\cite{Bishop98a} that CCM is more efficient 
computationally in the sense that, as the desired error goes to zero, the 
amount of computation required for CCM becomes negligible compared to a 
pure Cauchy computation.

We further apply our algorithms to the study of the propagation of a linearized
 Teukolsky gravitational wave~\cite{Nakao01, Kidder01a, Bonazzola:2003dm, 
Fiske05} and we compare the CCE and Zerilli wavesignal. 
For the case of a large extraction radius, we find that both methods give 
very good results and we demonstrate convergence to the analytical waveform 
of the Teukolsky solution for the CCE news. 
We show that the CCE waveform does not depend upon the extraction radius, 
which is a major advantage of the CCE method. A small extraction radius has no 
effect on the CCE waveform but it introduces errors in the Zerilli waveform.

In Sec.~\ref{sec:notation} we outline some general background and notation. 
In Sec.~\ref{sec:cce} we detail the transformation from the Cauchy slice to 
the characteristic slice. In Sec.~\ref{sec:bondi} we give a brief summary of 
the characteristic evolution code. In Sec.~\ref{sec:news} we describe how
the news is computed at $\scri^+$. Finally, in Sec.~\ref{sec:tests} we
give two sets of tests in full 3D numerical relativity to validate the
accuracy, convergence and robustness of the CCE algorithm. 
The first set contains four non-radiative tests, with no gravitational wave 
content. The second one is a radiative three-dimensional test. The results 
show that CCE is valid and accurate in both non-radiative and truly radiative situations.

\section{Notation, geometry and metrics}
\label{sec:notation}

The implementation of CCE described here follows previous descriptions
of Cauchy-characteristic extraction and matching in the
literature. Much of the work has been presented earlier~\cite{Bishop96,
Bishop98b,Winicour98,Winicour05, Szilagyi00}. Here we briefly
outline the notation, geometry and metrics used.

The geometry is described by two separate foliations, neither of which
cover the entire spacetime. The Cauchy foliation is described using a
standard $3+1$ ADM type metric~\cite{Arnowitt62},
\begin{equation}
  \label{eq:metric_adm}
  ds^2 = - (\alpha^2 - \beta_i \beta^i) dt^2 + 2 \beta_i dt dx^i +
  \gamma_{ij} dx^i dx^j.
\end{equation}
Many different formulations can be used, given initial and boundary
data, to evolve the 3-metric $\gamma_{ij}$. In what follows we shall
either consider the BSSN formulation~\cite{Shibata95,Baumgarte99} as
implemented in~\cite{Alcubierre02a} or the generalized harmonic
formulation~\cite{Wald84} as implemented in the Abigel
code~\cite{Szilagyi02a}. In both cases all interaction between the
Cauchy and characteristic foliations will be performed in terms of the
ADM metric, Eq.~\eqref{eq:metric_adm}.

The Cauchy slice does not extend to asymptotic infinity. Instead an
artificial boundary is placed at $|x^i| = L^i$. Within this artificial
boundary a {\it world-tube} $\Gamma$ is constructed such that its
intersection with any $t=\textrm{const.}$ Cauchy slice is defined as a
Cartesian sphere $x^2+y^2+z^2=R^2$, with angular coordinates labeled
by $\tilde{y}^{\tilde{A}}, \tilde{A} = (2,3)$. The world-tube $\Gamma$
is then used as the inner boundary of a characteristic foliation which
uses the standard Bondi-Sachs metric~\cite{Bondi62, Sachs62}
\begin{equation}
  \label{eq:metric_bondi}
  ds^2 = - \left( e^{2\beta} \frac{V}{r} - r^2 h_{AB} U^A U^B \right)
  du^2 - 2 e^{2\beta} du dr - 2 r^2 h_{AB} U^B du dy^A + r^2 h_{AB} dy^A
  dy^B. 
\end{equation}
Here $u$ labels the outgoing null hypersurfaces, $y^A=\tilde{y}^{\tilde{A}}$
the angular coordinates (the null rays emanating from the world-tube), 
and $r$ is a radial surface area distance. 
The angular metric $h_{AB}$ obeys the condition
\begin{equation}
  \label{eq:metric_det_condition}
  \det (h_{AB}) = \det (q_{AB}), 
\end{equation}
where $q_{AB}$ is the unit sphere metric. This coordinate system
consistently covers the world-tube $\Gamma$ and the exterior spacetime
as long as it is free of caustics.

The free variables in the Bondi-Sachs metric are then $V, \beta, U^A$ and
$h_{AB}$. The physical interpretation of these variables is that
$h_{AB}$ contains the 2 radiative degrees of freedom, $e^{2\beta}$ measures
the expansion of the nullcone, and $V/r$ is the counterpart of the Newtonian
gravitational potential. The 2+1 decomposition of the intrinsic metric on 
the $r=\textrm{const.}$ world-tube
\begin{equation}
  \label{eq:metric_bondi_intrinsic}
  \gamma_{ij} dy^i dy^j = - e^{2 \beta} \frac{V}{r} du^2
  + r^2 h_{AB} (dy^A - U^A du) (dy^B - U^B du)
\end{equation}
identifies $r^2 h_{AB}$ as the intrinsic 2-metric of the $u$ foliation, 
$-U^A$ as the shift vector and $e^{2 \beta} V / r$ as the square of the 
lapse.

\section{The CCE algorithm}
\label{sec:cce}

The crucial task of the CCE algorithm is to take Cauchy data given in
the ADM form Eq.~\eqref{eq:metric_adm} in a neighborhood of the
world-tube $\Gamma$ and to transform it into boundary data for the
Bondi-Sachs metric Eq.~\eqref{eq:metric_bondi}. Then the Bondi code can
use the hypersurface equations to evolve the appropriate quantities
out to $\scri^+$ so that the gravitational news may be extracted
there.  Much of the present version of the CCE algorithm has been
presented in earlier work~\cite{Bishop96,Bishop98b,Winicour98, 
Winicour05, Szilagyi00}, so in this section we will give a brief description, 
highlighting a few new features.

In section~\ref{sec:notation} the world-tube was defined as a
Cartesian sphere $x^2+y^2+z^2=R^2$, with angular coordinates labeled
by $\tilde{y}^{\tilde{A}}, \tilde{A} = (2,3)$. In addition to the
angular coordinates, we set $u=t$ on the world-tube and choose the
fourth coordinate, $\lambda$ to be the affine parameter along the
radial direction, with $\lambda_{|\Gamma} = 0$.  The characteristic
cones are constructed such that $\lambda$ is the future oriented,
outgoing null direction normal to the foliation of $\Gamma$. (In order 
to avoid a singular Jacobian for the Cauchy-characteristic coordinate
transformation, we require that the world-tube $\Gamma$ be timelike.)
The affine parameter $\lambda$ is used because the world-tube is 
not constructed to be a surface of constant Bondi $r$.

The choice of the angular coordinates is determined
following~\cite{Gomez97} by the use of two stereographic patches. The
use of two patches avoids numerical difficulties in taking derivatives
near the poles when standard spherical coordinates are used. The use of only
two patches may not give the most accurate numerical results; various
ways of discretizing the 2-sphere on multiple patches are discussed
in~\cite{Thornburg2004:multipatch-BH-excision}. The two patches are
centered around the North and South poles with the stereographic
coordinates related to the usual spherical coordinates $(\theta, \phi)$ by
\begin{equation}
  \label{eq:xi}
  \xi_{\rm North} = \sqrt{ \frac{1 - \cos\theta} 
    {1 + \cos\theta}} e^{i\phi},
  \quad
  \xi_{\rm South} = \sqrt{ \frac{1 + \cos\theta} 
    {1 - \cos\theta}} e^{-i\phi},
\end{equation}
and $\xi = \tilde y^2 + i \tilde y^3 = q + i\, p,$ where $i =
\sqrt{-1}$ .  We also introduce the complex vector on the sphere $q^A
= (P/2) ( \delta_2^A + i \delta_3^A )$ and its co-vector $q_A = (2/P)
( \delta^2_A + i \delta^3_A )$, with $P = 1 + \xi \bar \xi = 1 + q^2 +
p^2$.  The orthogonality condition $\bar q_A q^A = 2$ is satisfied by
construction.  The unit sphere metric corresponding to these
coordinates is
\begin{equation}
  \label{eq:metric_unitsphere}
  q_{AB} = \frac{1}{2} \left( q_A \bar q_B + \bar q_A q_B \right)
  = \frac{4} {P^{2}} 
       \left [
              \begin{array}{cc}
              1 & 0
              \\
              0 & 1
              \end{array}
       \right ].
\end{equation}

The angular subspace in the Bondi code is treated by use of the {\em
  eth} formalism and spin-weighted quantities. (See~\cite{Gomez97,Bishop97b} for details.)

The CCE algorithm starts by interpolating the Cauchy metric
$\gamma_{ij}$, lapse $\alpha$ and shift $\beta^i$ and their spatial
derivatives onto the world-tube.  Time derivatives are computed via
backwards finite-differencing done along $\Gamma$, e.g., at $t=t_N$ we
write
\begin{equation}
  \label{eq:timederiv1}
  (\partial_t F)_{[N]} = \frac{1}{2 \Delta t}
  \bigg( 3 \, F_{[N]} - 4 F_{[N-1]} + F_{[N-2]} \bigg)
  + {\cal O}\left((\Delta t)^2\right).
\end{equation}
Knowledge of the Cauchy metric and its 4-derivative is enough to
compute the affine metric $\tilde \eta^{\tilde \alpha \tilde \beta}$
as a Taylor series expansion around the world-tube
\begin{equation}
  \label{eq:taylorexp}
  \tilde \eta^{\tilde \alpha \tilde \beta}
  = 
  \tilde \eta^{\tilde \alpha \tilde \beta}|_\Gamma
  + \lambda
  \tilde \eta^{\tilde \alpha \tilde \beta}_{,\lambda}|_\Gamma
  + {\cal O}(\lambda^2).
\end{equation}
Next a second null coordinate system (the Bondi-Sachs system) is introduced
\begin{equation}
  \label{eq:bondicoord}
  y^\alpha = (y^1, y^A, y^4), \quad \mbox{where}
  \quad y^A = \tilde y^{\tilde A},  \quad
  y^1 = \tilde y^1 = u.
\end{equation}
The fourth coordinate $y^4=r$ is a surface area coordinate, defined by
\begin{equation}
  \label{eq:rofeta}
  r  =  \left( \frac{\det(\tilde{\eta}_{\tilde{A}\tilde{B}})}
    {\det(q_{AB})} \right) 
  ^{\frac{1}{4}} = \frac{P}{2} \;
  {\det(\tilde{\eta}_{\tilde{A}\tilde{B}})}^{\frac{1}{4}}. 
\end{equation}
The Bondi-Sachs metric on the extraction world-tube can be computed as
\begin{equation}
  \label{eq:metric_bondi_worldtube}
  \eta^{\mu\nu}_{|\Gamma} =
  \bigg( \frac{ \partial y ^{\mu} }{\partial \tilde y^{\tilde \alpha}} \, 
  \frac{ \partial y ^{\nu} }{\partial \tilde y^{\tilde \beta}} \, 
  \tilde \eta^{\tilde \alpha\tilde \beta}\bigg)_{|\Gamma}.
\end{equation}
Note that the metric on the sphere is unchanged by this coordinate
transformation, i.e., $\eta^{AB} = \tilde \eta^{\tilde A \tilde B}$.
Therefore one only needs to work with the Jacobian components that
correspond to derivatives of $r$.

In terms of $q^A, \bar q^A$ the two dimensional metric $\eta^{AB}$ can
be encoded into the metric functions
\begin{equation}
  \label{eq:jandkencode}
  J \equiv \frac{1}{2}q^A q^B h_{AB}, \quad 
  K \equiv \frac{1}{2}q^A \bar q^B h_{AB} \, .
\end{equation}
The determinant condition Eq.~\eqref{eq:metric_det_condition}
translates into
\begin{equation}
  \label{eq:Kconstr}
  K^2 = 1 + J \bar J .
\end{equation}
With $h_{AB}$ symmetric and of fixed determinant, there are only two
degrees of freedom in the angular metric that are encoded into the
complex function $J$.

From Eq.~\eqref{eq:metric_bondi} we have 
\begin{equation}
  \label{eq:redshift}
  \beta = - \frac{1}{2} \log( - \eta^{rr} ) =  \frac{1}{2} \log(
  r_{,\lambda} ). 
\end{equation}
This quantity is a measure of the expansion of the light rays as
they propagate outwards. The CCE and the Bondi codes have been
implemented with the assumption that $r_{,\lambda} > 0$ (or that
$\beta$ is real).

The radial-angular components $\eta^{rA}$ can be represented by
\begin{equation}
  \label{eq:radial-angular}
  U \equiv U^A q_A = \frac{\eta^{rA}}{\eta^{ru}};
\end{equation}
while the radial-radial component $\eta^{rr}$ is contained in
\begin{equation}
  \label{eq:radial-radial}
  \quad W \equiv \frac{V-r}{r^2} \; .
\end{equation}

In addition to these quantities, in~\cite{Gomez01} the auxiliary
variables
\begin{eqnarray}
  \label{eq:bondiauxiliary}
  \nu &\equiv& \bar \eth J = \frac{1}{2} h_{AB,C} q^A q^B \bar q^C ,\\
  k &\equiv& \eth K = \frac{1}{2} h_{AB,C} q^A \bar q^B q^C + 2 \xi K ,\\
  B &\equiv& \eth \beta = \beta_{,A} q^A,
\end{eqnarray}
have been introduced to eliminate the need to explicitly use second
angular derivatives in the Bondi evolution code.  The required
boundary data is $J, \beta, U, \partial_r U, W, \nu, k,$ and $B$. 
(See Sec.~\ref{sec:bondi}.)  Notice that once the Bondi-Sachs metric
is known on the world-tube one can only obtain $J, \beta, U, W$.  In
order to provide the rest of the necessary boundary data we need the
radial derivative of the Bondi-Sachs metric
\begin{equation}
  \label{eq:eta1}
  \left(\partial_\lambda \eta^{\mu\nu}\right)_{|\Gamma} = 
  \bigg( \frac{ \partial^2 y ^{\mu} }{\partial \lambda \partial \tilde
    y^{\tilde \alpha}} \,  
  \frac{ \partial y ^{\nu} }{\partial \tilde y^{\tilde \beta}} \, 
  \tilde \eta^{\tilde \alpha\tilde \beta}
  +      \frac{ \partial y ^{\mu} }{\partial \tilde y^{\tilde \alpha}} \, 
  \frac{ \partial^2 y ^{\nu} }{\partial \lambda \partial \tilde
    y^{\tilde \beta}} \,  
  \tilde \eta^{\tilde \alpha\tilde \beta}
  +      \frac{ \partial y ^{\mu} }{\partial \tilde y^{\tilde \alpha}} \, 
  \frac{ \partial y ^{\nu} }{\partial \tilde y^{\tilde \beta}} \, 
  \tilde \eta^{\tilde \alpha\tilde \beta}_{,\lambda}\bigg)_{|\Gamma}.
\end{equation}
With $\tilde \eta^{\tilde \alpha\tilde \beta}_{,\lambda}$ already
known, the only non-trivial parts in Eq.(\ref{eq:eta1}) are the
Jacobian terms
\begin{equation}
  \label{eq:jacobian}
  r_{,\lambda \tilde \alpha} = \frac{ \partial^2 y ^{1} }{\partial
    \lambda \partial \tilde y^{\tilde \alpha}} \, . 
\end{equation}
These depend on the second derivatives of the Cauchy metric. In order
to avoid possible numerical problems caused by interpolating
second derivatives onto the world-tube, we calculate $r_{,\lambda
  \tilde A}$ by taking centered derivatives of $r_{,\lambda}$ on the
world-tube and we calculate $r_{,\lambda u}$ by backwards differencing
in time along the world-tube.  The remaining term $r_{,\lambda
  \lambda}$ is calculated using the identity
\begin{equation}
  \label{r-lambda}
  \beta_{,\lambda} = - \frac{r_{,\lambda\lambda}}{2\,r_{,\lambda}},
\end{equation}
and the characteristic equation
\begin{equation}
  \label{eq:beta_r}
  \beta_{,r} = \frac{r}{8} \left( J_{,r} \bar J_{,r} - (K_{,r})^2
  \right) \, . 
\end{equation}
(The right hand side of Eq.~\eqref{eq:beta_r} can be computed in terms
of $J_{,r} = J_{,\lambda} / r_{,\lambda} $, which, in turn, can be
computed in terms of already known quantities.)
 
Knowledge of the radial derivative of the Bondi metric is important
not only for obtaining $\left(\partial_r U\right)_{|\Gamma}$ but also
because the grid-structure of the Bondi code is based on the radial
coordinate $r$, and the extraction world-tube will not, in general,
coincide with any of these radial grid-points.  We need to use,
therefore, Taylor series expansions to fill the Bondi gridpoints
surrounding $\Gamma$ with the necessary boundary data.  We write,
e.g.,
\begin{equation}
  \label{eq:beta-expansion}
  \beta = \beta_{|\Gamma} + \lambda \beta_{,\lambda} + {\cal O}(\lambda^2).
\end{equation}

Another problem is the need to provide the auxiliary angular
variables, $B,k$ and $\nu$ on the world-tube.  These have been defined
as the $\eth$-derivatives of Bondi fields in the $y^\alpha$ frame,
while taking angular derivatives on the world-tube amounts to
computing $\eth$ derivatives in the $\tilde y^{\tilde \alpha}$ frame.
The correction term between the two frames is
\begin{equation}
  \label{eq:f-correction}
  \eth F =\tilde\eth F-\frac{F_{,\lambda}}{r_{,\lambda}}\tilde \eth r.
\end{equation}
The quantities $J, \tilde \eth J$ and $J_{,\lambda}$ are known from
the interpolated Cauchy data, while $J_{,\lambda\lambda}$ is not
computed in CCE.  Thus one can compute $\eth J$ but not $\eth
J_{,\lambda}$.  As a consequence we cannot use a simple Taylor series
expansion to place $\nu$ on the Bondi grid to ${\cal O}(\lambda^2)$
accuracy. The solution to this problem is to first place $J$ on
Bondi grid-points surrounding the world-tube, then compute $\eth J$ on
those points (i.e., compute angular derivatives in the Bondi-Sachs frame),
and then calculate $\partial_\lambda \eth J$ on the world-tube by use
of the neighboring Bondi grid values of $\eth J$.  With $\nu = \eth
J$ and its radial derivative known on the world-tube, we can then use
the standard ${\cal O}(\lambda^2)$ expansion to provide $\nu$ at Bondi
gridpoints surrounding the world-tube.  This way we make maximal use
of the Cauchy data as interpolated onto the world-tube and minimal use
of the finite difference $\eth$ algorithm that has its own
discontinuous ${\cal O}(\Delta^2)$ error at the edges of the
stereographic patches.  A similar approach is used for $k = \eth K$
and $B = \eth \beta$.

\section{The Bondi code and the News algorithm}
\label{sec:bondinews}

\subsection{The Bondi code}
\label{sec:bondi}

The inner workings of the Bondi code had been described in detail
elsewhere (see~\cite{Bishop96, Gomez97b, Gomez97a, Bishop97b,
Lehner98, Winicour98, Winicour05, Lehner99a, Gomez01}) so here
we give only a brief overview of the algorithm.  As already stated in
Sec.~\ref{sec:cce}, the variables of the code are $J, \beta, U, W$ as
well as $\nu, k, B$.  Out of these the equation for $J$ is the only
one to contain a time derivative. For this reason $J$ is updated via
an evolution stencil that involves two time-levels.  The rest of the
variables are integrated radially from the world-tube out to
$\scri^+$.  The integration constants are set by CCE.  All of these
radial integration equations are first differential order in $r$
except for the $U$ equation which contains $U_{,rr}$.  For this reason
integrating $U$ requires two constants, which explains the need to
provide, at the world-tube, $U$ as well as $U_{,r}$.

\subsection{The News algorithm}
\label{sec:news}

The calculation of the Bondi news function on $\scri^+$ is based on
the algorithm developed in~\cite{Bishop97b} with the modifications
introduced in~\cite{Zlochower02,Zlochower03} (an alternative calculation for the
news was recently introduced in~\cite{Bishop03}). Here we present
an overview of the algorithm. 

The Bondi-Sachs metric Eq.~\eqref{eq:metric_bondi} in a neighborhood of
$\scri^+$ in $(u, x^A, l=1/r)$ coordinates (after multiplying by a
conformal factor $l^2$) has the form
\begin{equation}
  \label{eqn:codemetscri}
  l^2\,ds^2 = {\cal O}(l^2) du^2 + 2 e^{2 \beta}du\,dl -2
  h_{AB}U^B du\,dx^A+h_{AB} dx^A\,dx^B,
\end{equation}
where the metric variables $\beta$, $U^A$, and $h_{AB}$ have the
asymptotic expansions $\beta = H + {\cal O}(l^2)$, $U^A = L^A + {\cal
  O}(l)$, and $h_{AB} = H_{AB} + l\,c_{AB} + {\cal O}(l^2)$.  We can
always find coordinates $(u_B, q_B, p_B, l_B)$ (hereafter referred to
as `inertial' coordinates), and an associated conformal metric
${ds_B}^2 = \omega^2 ds^2$ ($\omega > 0$), such that (i)
$\left(\frac{\partial}{\partial u_B}\right)$ is null and affine and
points along the null generators of $\scri^+$, (ii) $l_B = \omega\, l
+ {\cal O}(l^2)$, (iii) the conformal metric in the subspace
$(u_B=\textrm{const.}, l_B=\textrm{const.})$ is the unit sphere metric
on $\scri^+$.  We fix a null-tetrad on $\scri^+$ by choosing $n^a$ to
be affine and to point along the null generators of $\scri^+$ and
by choosing $m^{(a}\bar m^{b)}$ to be the unit-sphere covariant metric in
the 2-dimensional angular subspace.  In inertial coordinates ($u_B,
q_B, p_B, l_B$) on $\scri^+$ the tetrad has the form
\begin{eqnarray}
  \label{eq:tetrad}
  \tilde n^a &=& (1,0,0,0) \\
  \tilde l^a &=& (0,0,0,1) \\
  \tilde m^a &=& (0,\tilde P/2, i \tilde P/2,0),
\end{eqnarray}
where the tilde denotes a quantity defined with respect to inertial
observers. Note that ${\tilde m}^{(A} {\bar{\tilde m}}{}^{B)} = q^{AB}$ as required.  We
define a complex vector $F^a$, analogous to $\tilde m^a$ (i.e.
$F^{(A}\bar F^{B)}=H^{AB}$), adapted to the coordinates used in the
characteristic evolution code via
\begin{equation}
  \label{eq:fa-coords}
  F^a = (0,F^A,0),  
\end{equation}
where 
\begin{equation}
  \label{eq:fa}
  F^A  = q^A \sqrt{\frac{K_0+1}{2}} - J_0\bar q^A \sqrt{\frac{1}{2 (K_0+1)}},
\end{equation}
where $q^A$ is the dyad defined in Sec.~\ref{sec:cce}, $J_0=q^A q^B
H_{AB}/2$, and $K_0 = q^A \bar q^B H_{AB}/2$.  $F^a$ and $\tilde m^a$ are
related by
\begin{equation}
  \label{eq:mafa}
  \tilde m^a = e^{-i \delta} \omega^{-1} F^a + \gamma \tilde n^a.
\end{equation}
The $\gamma \tilde n^a$ term will not enter the news calculation.

To calculate the Bondi news one needs to evolve two scalar quantities
$\delta$, the phase factor in Eq.~\eqref{eq:mafa}, and the conformal
factor $\omega$, as well the relations $\xi(\xi_B)$ between the
angular coordinates used in the characteristic evolution and inertial
angular coordinates, and $u_B(u,\xi_B)$ between the inertial time
slicing and the time slicing of the characteristic evolution code.
Then  $\delta$, $\xi(\xi_B)$, and $u_B(u,\xi_B)$ are evolved using
the following ODE's along the null generators of $\scri^+$
\begin{eqnarray}
  \label{eq:newsevolveode}
  \frac{d\delta}{du} &=& \frac{1}{2}\Im\left(\frac{{\bar{J}}_{0,u}J_0}{K_0+1}
    + 
    \frac{J_0\left(U_0 \bar\eth\bar J_0 + \bar U_0 \eth \bar J_0\right)}{2(K_0+1)}
    + 
    J_0 \bar \eth \bar U_0 + K_0 \bar \eth U_0  + 2 U_0 \bar \xi \right)\\
  \frac{d \xi}{du} &=& \frac{1}{2}(1+\xi \bar \xi)U_0 \\
  \frac{d u_B}{du} &=& \omega e^{2 H},
\end{eqnarray}
where $U_0=q_A\,L^A$. Also $\omega$ is evolved using the PDE
\begin{equation}  
  \label{eq:newsevolvepde}
  \partial_u \log \omega = - \Re\left(\bar U\, \eth \log \omega +
    \frac{1}{2} \eth \bar U\right)
\end{equation}  

The Bondi news function up to a phase factor of $e^{-2 i \delta}$ is
given by
\begin{equation}
  \label{eq:News}
  N = \frac{1}{2} \omega^{-2} e^{-2 H} F^A F^B \biggl(
  \left(\partial_u + \LIE{L}\right) c_{AB}  - \frac{1}{2} c_{AB} D_C L^C + 2 \omega
  D_A [\omega^{-2} D_B(\omega e^{2 H})]\biggr),
\end{equation}
where $D_A$ is the covariant derivative with respect to $H_{AB}$.  The
Bondi news is calculated in three steps. First Eq.~\eqref{eq:News} is
evaluated, ignoring the $e^{-2i\delta}$ phase factor, as a function of
the evolution coordinates $(u, \xi)$. Then using the relation
$\xi(\xi_B)$, the news is interpolated onto a fixed inertial angular
grid (i.e.\ $N(u,\xi_B)$) and multiplied by the phase factor
$e^{-2i\delta}$ (which is only known on the inertial grid). Finally,
in post-processing, the news is interpolated in time onto fixed
inertial time slices (i.e.\ $N(u_B, \xi_B)$). Once the news is
obtained in inertial coordinates it can be decomposed into
spin-weighted spherical harmonics.

\section{Tests}
\label{sec:tests}

We apply the algorithm described above for two sets of tests: non-radiative 
spacetimes and radiative spacetimes.
In the first, non-radiative set of tests (section~\ref{sec:tests_non}), 
we analyze Minkowsy and Schwazchild spacetimes. 
The Minkowski (Sec.~\ref{sec:tests_flat}) or small perturbation of 
Minkowski (Sec.~\ref{sec:tests_rand}) tests are used
to show the stability of the code and the errors due to transforming
between the different coordinate systems and sets of variables. 

The Schwazchild tests describe a static spherically symmetric black hole 
in a centered frame (Sec.~\ref{sec:tests_bh_fix}) and in an oscillating 
frame (Sec.~\ref{sec:tests_bh_bbh}).
These tests indicate the accuracy of the code in non-trivial spacetimes.

The last test (Sec.~\ref{sec:tests_teuk}) is a standard radiative test: 
namely a linearized Teukolsky wave, and indicates that CCE is a valid and 
accurate method in truly radiative situations.

In the non-radiative cases, the extracted gravitational wave signal should 
vanish identically. As this is true uniformly for all points at $\scri^+$ we
simplify the computation for the norms of the gravitational wave
signal. When the results of the CCE are shown
the norm used is the simple $L_2$ norm over all grid points at
$\scri^+$, without any weighting by the area element. 
In the radiative case, we plot the real part of the extracted news 
waveform ($\partial_t h_+$), versus time.

The Zerilli Moncrief formalism (see~\cite{Zerilli70,Moncrief74,Camarda97c} 
for the implementation used here) approximates the spacetime as a linearized
perturbation of a spherically symmetric spacetime. Such perturbations
are decomposed in terms of spherical harmonics which are then
expressed in terms of some basis set. An example using the
Regge-Wheeler set is where the even parity metric perturbations are
decomposed as
\begin{equation}
   \label{eq:ZerilliEvenParity1}
   \left( h^{lm}_{\mu\nu} \right)^{(e)} = \left(
     \begin{array}{c c|c}
       e^{2a} H_0 Y^{lm} & H_1 Y^{lm} & \\
       H_1 Y^{lm} & e^{2b} H_2 Y^{lm} & H^{(e)}_A \nabla_c Y^{lm} \\
       \hline
       h^{(e)}_A \nabla_c Y^{lm} & & r^2 \left( K Y^{lm} \gamma_{cd} +
         G \nabla_d \nabla_c Y^{lm} \right)
     \end{array}
   \right).
\end{equation}
Once the basis functions such as $G,K,h^{(e)}_1,H_2$ and $b$ are found
from the decomposition, the even parity master function $Q^{+}$ can be
computed from
\begin{equation}
   \label{eq:ZerilliEvenParity2}
   Q^{+} = \sqrt{\frac{2(l+2)!}{(l-2)!}} \frac{r \left[ r \Lambda
       \left( K + \frac{1}{e^{2b}} \left( r \partial_r G - \frac{2}{r}
             h^{(e)}_1 \right) \right) + \frac{2r}{e^{4b}}\left( e^{2b}
             H_2 - e^b \partial_r \left(r e^b K\right) \right)
         \right]}{\Lambda \left[ r (\Lambda -2) +6M \right]}.
\end{equation}
Equivalent definitions in terms of other basis functions, alternative
conventions and expressions for the odd parity sector can be found
in~\cite{Nagar05}.

Using this notation where $Q^{\times}$ gives the odd parity master function, 
the asymptotic value for the wave forms is
   \begin{equation}
     \label{eq:Zerillih1}
     h_{+} - \text{i}h_{\times} = \frac{1}{\sqrt{2}r} \sum_{l, m}
     \left(Q^{+}_{lm} - \text{i} \int^t_{-\infty} Q^{\times}_{lm}(t')
       \text{d}t' \right) {}_{-2}Y^{lm}(\theta,\phi) + \mathcal{O}
     \left( \frac{1}{r^2} \right).
   \end{equation}
For the non-radiative cases, when we compute the wavesignal from the Zerilli 
approximation, we use the norm
\begin{eqnarray}
  \label{eq:Zerilli_norm}
  (\|h\|_2)^2 & = & \int (h_+ - i h_{\times}) (h_+ - i
                      h_{\times})^{\dagger} d \Omega \\
              & = & \frac{1}{2r^2} (Q^+_{\ell m} - i
                      q^{\times}_{\ell m})  (Q^+_{\ell ' m'} - i
                      q^{\times}_{\ell ' m'})^{\dagger} \int {}_{-2}Y^{\ell m}
                      {}_{-2}{Y^{\ell ' m '}}^{\dagger} d \Omega \nonumber \\
\label{eq:Zerilli_norm3}
              & = & \frac{1}{2r^2} \delta^{\ell \ell '} 
                                     \delta^{ m m'} 
                      (Q^+_{\ell m} - i q^{\times}_{\ell m})  
                      (Q^+_{\ell ' m'} - i q^{\times}_{\ell ' m'})^{\dagger},
\end{eqnarray}
where $d \Omega$ is the element of solid angle. The last equality Eq.~\eqref{eq:Zerilli_norm3} follows from the orthogonality of the
spin-weighted tensor spherical harmonics. We expect $h_+ - i h_\times$ to 
vanish for all angles so this norm should be identically zero. 
In what follows we shall look at the spherical harmonics up to $\ell = m
= 7$ although in practice there is little contribution from the higher modes.

For the Teukolsky wave used here we expect all terms except for the
$Q^{+}_{2,0}$ term in Eq.~\eqref{eq:Zerillih1} to be negligible asymptotically, 
which leads to a simple but tedious computation of the wavesignal.

\subsection{Non-radiative Spacetime Tests}
\label{sec:tests_non}

\subsubsection{Minkowski Flat Spacetime}
\label{sec:tests_flat}

Minkowski space in standard coordinates is evolved in the harmonic
Abigel code, with the metric given analytically on all Cauchy slices. 
The domain extends between $x^i \in [-10,10]$, the extraction world-tube 
is placed at $r=7$, and the simulation is run until $t=10$. The
coarsest simulation used $50^3$ points for the Cauchy grid and $35^2
\times 31$ points for the characteristic grid.  Simulations to test
convergence scaled all grids by factors of $2$. For this test,
extraction using the Zerilli method gives results that are identically
zero, as expected.

\begin{figure}[htbp]
  \centering
  \includegraphics[width=7.5cm]{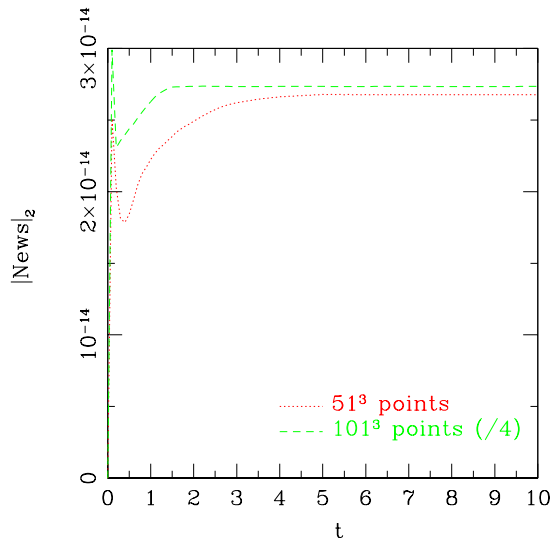}
  \caption{The $L_2$ norm of the CCE news for Minkowski space in standard
    coordinates. This indicates the level of numerical noise
    introduced by transforming variables on the world-tube. In this
    case the noise is linearly dependent on the number of points on
    the extraction world-tube. It is, however, extremely small.}
  \label{fig:News_mink}
\end{figure}

This test indicates the level of numerical round off error in
transforming from the Cauchy variables to the Bondi variables on the
world-tube. As shown in Fig.~\ref{fig:News_mink}, the extracted news
is small in all cases but the level of truncation error increases
linearly with the number of points on the extraction world-tube.
Since the construction of the boundary data for the
CCE code involves derivatives of the Cauchy metric,
this sensitivity to noise in the Cauchy data is
expected.  However, given the extremely low amplitude of the noise,
it is unlikely that this will cause problems in practical simulations.

\subsubsection{Random Perturbations around Minkowski Flat Spacetime}
\label{sec:tests_rand}

This test is in the spirit of the robust stability tests
of~\cite{Szilagyi00a,Szilagyi02b,Alcubierre2003:mexico-I}. The stable
evolution and extraction of white noise initial data with no frequency
dependent growth of the wavesignal is a good indication that the
combined evolution and extraction codes are stable.  The domain extends 
between $x^i \in [-10,10]$, the extraction world-tube is placed at $r=7$,
and the simulation is run until $t=100$. The coarsest Cauchy grid has
$51^3$ points and the coarsest characteristic grid $35^2 \times 31$
points. 

\begin{figure}[htbp]
  \centering
  \includegraphics[width=7.5cm]{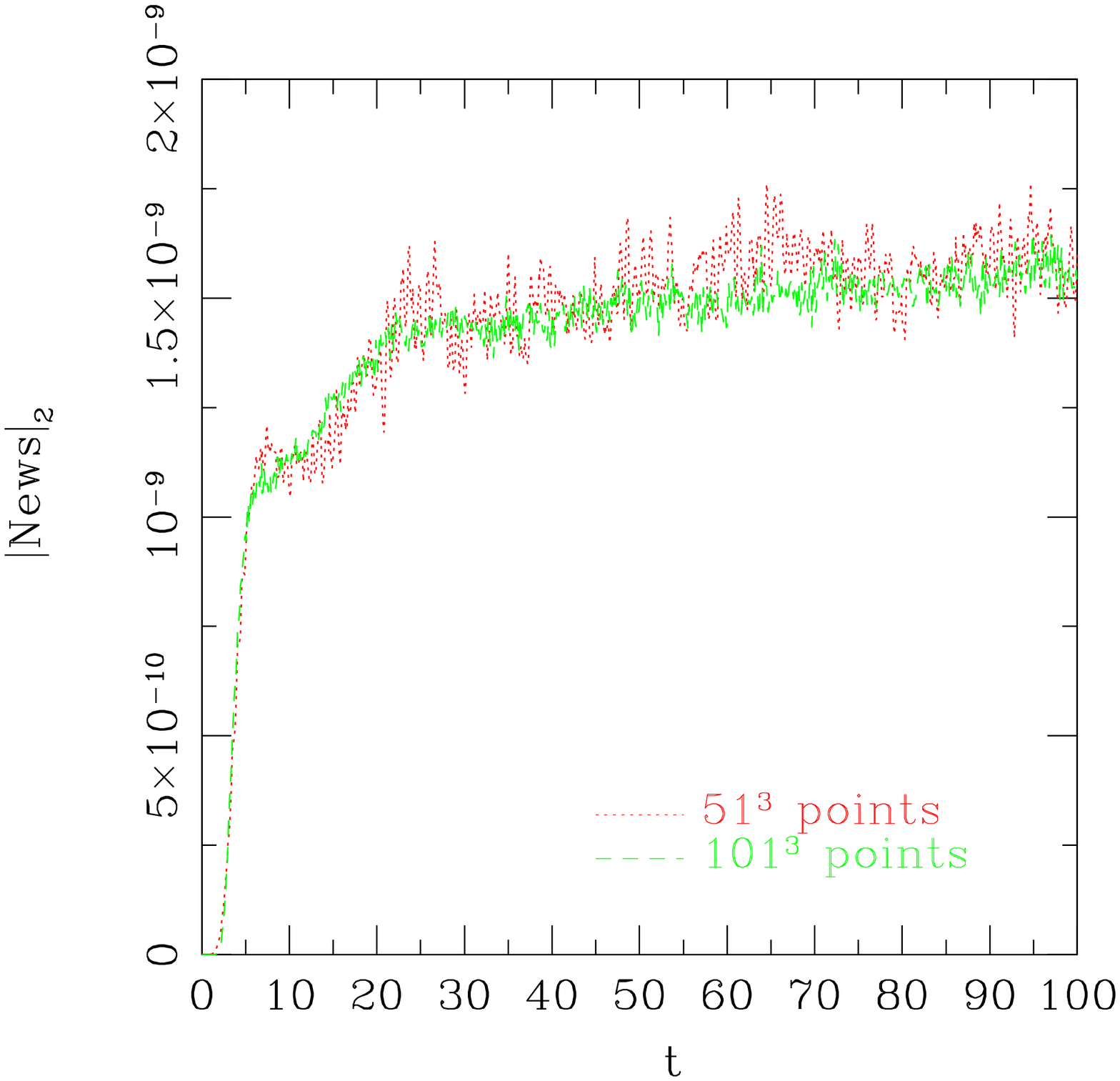}
  \includegraphics[width=7.5cm]{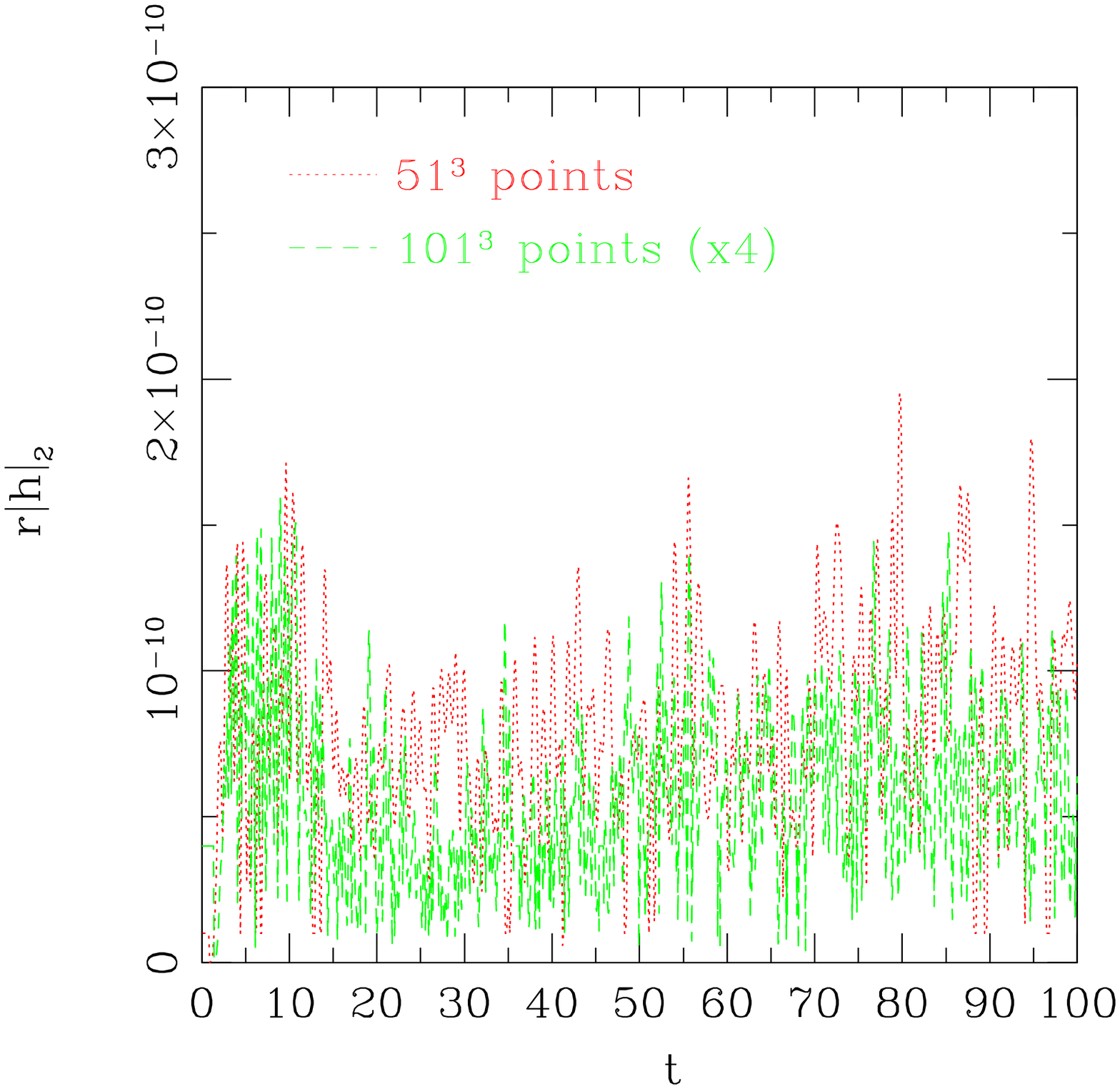}
  \caption{The $L_2$ norm of the CCE news for random perturbations of
    Minkowski where the Cauchy slice is evolved using the Abigel code.
    The news displays very slow growth that is independent of the
    frequency of the initial data. This is a good indication that the
    CCE method is stable. The right hand figure shows that the norm of the 
    waveform extracted from the Cauchy grid using the Zerilli method is 
    about one order of magnitude smaller.}
  \label{fig:News_rand}
\end{figure}

The results in Fig.~\ref{fig:News_rand} show that the error is independent 
of the resolution of both Cauchy and characteristic grids. 
This is strong evidence that the CCE code is stable against small 
perturbations that in a practical run would be induced by numerical error. 
The wavesignal from the Zerilli extraction is about an order of magnitude 
smaller. This is to be expected as the Zerilli extraction approach uses 
the field values. The error in the CCE news has a jump at the beginning. 
This arises because on the initial characteristic data is set to zero so 
that it takes some time for the noise to build up on the outgoing null cone.

\subsubsection{Schwarzchild Black Hole in a "centered" frame}
\label{sec:tests_bh_fix}

To test a simple black hole spacetime we use a Schwarzschild black hole 
in ingoing Eddington-Finklestein coordinates $(\hat{t}, {\hat{x}}^i)$, 
with the line element
\begin{equation}
  \label{eq:ds_ef}
  ds^2 = -\left(1-\frac{2M}{\hat{r}}\right)\, d\hat{t}^2
         + \left(\frac{4M}{\hat{r}}\right)\, d\hat{t}\, d\hat{r}
         + \left(1+\frac{2M}{\hat{r}}\right)\, d\hat{r}^2 +
         \hat{r}^2\, d\Omega^2.
\end{equation}
This is manifestly static in these coordinates. The spacetime is
evolved using the BSSN code and excision methods described
in~\cite{Alcubierre00a}. The coarsest Cauchy grid has $29^3$ points
with $\Delta x^i = 0.4M$, whilst the characteristic grid has $35^2
\times 31$ points. The world-tube is at $r=7M$. In the Cauchy
evolution domain octant symmetry is used.

\begin{figure}[htbp]
  \centering
  \includegraphics[width=7.5cm]{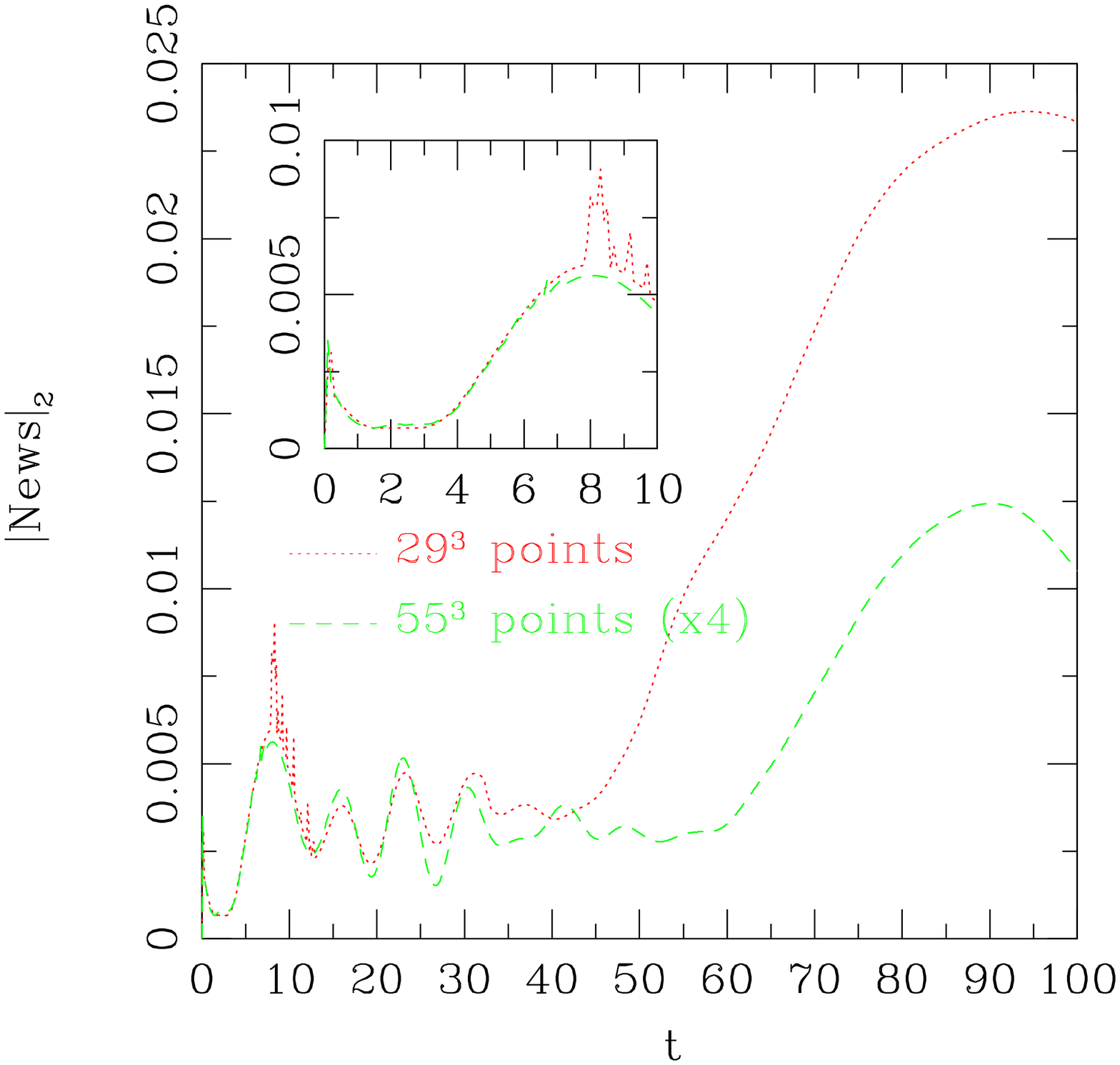}
  \includegraphics[width=7.5cm]{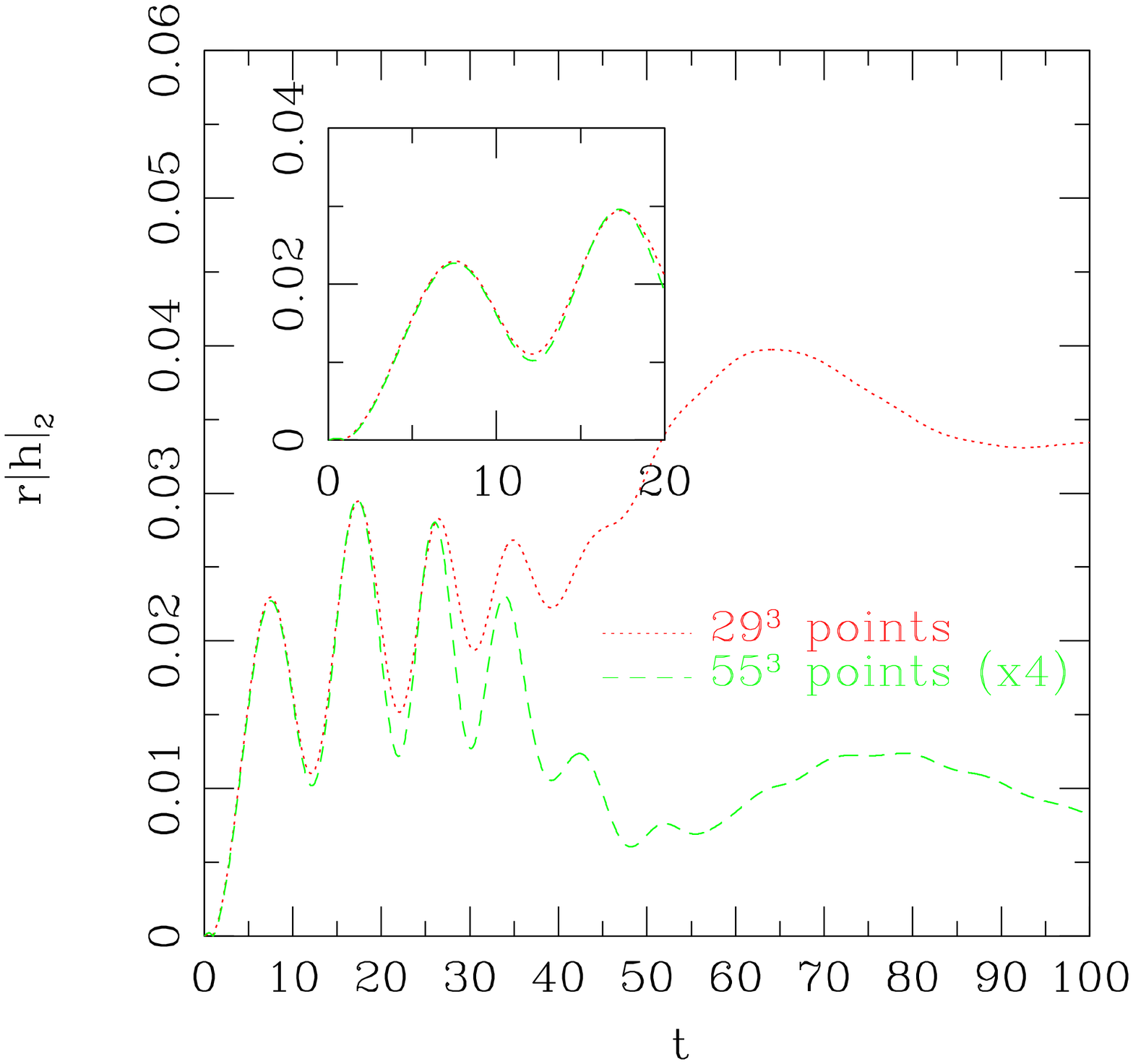}
  \caption{The scaled $L_2$ norm of the CCE news for a static 
    spherically symmetric black hole where the Cauchy slice is evolved using 
    the BSSN formalism. At early times,
    when the world-tube is causally disconnected from
    the outer boundary of the Cauchy slice, the CCE news converges to zero
    at second order as it should. Later, when the outer boundary is causally
    connected to the world-tube, deviations from second order
    convergence appear, indicating the effect of the Cauchy boundary
    conditions on the extracted wavesignal. The left panel shows the
    CCE news extracted at $\scri^+$, whilst the right panel shows the
    ``norm'' of the wavesignal extracted by the Zerilli method at $r=7$.}
  \label{fig:News_ef}
\end{figure}

The evolution is only performed for a short time (to $t=100M$). 
Over this timescale we see second order convergence for
the CCE news until $t \approx 7M$ and second order convergence in the
waveform from Zerilli extraction until $t \approx 20M$, as seen in
Fig.~\ref{fig:News_ef}. By varying the location of the world-tube or
the Zerilli extraction sphere we can see that the errors come from a
variety of locations. 

The difference between the times at which the two extraction methods lose 
convergence is probably due to the greater differencing error seen in the 
Zerilli extraction. At early times (where finite differencing error 
dominates) the absolute value of the error is larger for Zerilli 
extraction by a factor $\approx 4$, as seen in Fig.~\ref{fig:News_ef}.
At late times (where outer Cauchy boundary errors dominate) the error for 
Zerilli extraction is of the same order as for the CCE extraction.

The non-convergent errors are probably caused by the boundary
conditions on the Cauchy grid which do not satisfy the constraints. As
in~\cite{Alcubierre00a} we are simply applying Sommerfeld type
boundary conditions to all fields.  This condition does not {\em a
  priori} satisfy the constraints and is not known to be well-posed.
Thus we might expect errors to be induced by the use of these boundary
conditions. It is likely that constraint satisfying boundary
conditions or Cauchy-characteristic matching would solve or at least
greatly reduce this problem. Due to the complicated pattern of the
reflected errors arising from a boundary condition on a cubical surface,
we have been unable to determine exactly how this error depends on the 
location of the outer Cauchy boundary. 

\subsubsection{Schwarzchild Black Hole in an oscillating frame}
\label{sec:tests_bh_bbh}

We use the line element for the standard Schwarzschild black hole in ingoing 
Eddington-Finklestein coordinates, given in Eq.~\eqref{eq:ds_ef}. The moving
coordinate frame $(t, x^i)$ is given by
\begin{equation}
  \label{eq:ef_xform}
  t = \hat{t}, \, x^i = {\hat{x}}^i + B^i b(t),
\end{equation}
where $B^i$ are parameters specifying the velocity and $b(t)$ is
a simple periodic function turned on after some time $t>t_0$; here we use 
\begin{equation}
  \label{eq:ef_bfn}
  b(t) = \left\{
            \begin{array}{ll}
                0,                           &  t \leq t_0   \\
                \sin (\omega (t - t_0))^3,   &  t > t_0 .
            \end{array}
         \right . 
\end{equation}
For the test shown here we set $B^x = 0.2, B^y = 0.5, B^z = 0.3$ for
the velocity and $\omega = 2 \pi \times 0.05, t_0 = 0.5$ for the
function $b$.

The metric is given analytically on the Cauchy grid to avoid any
boundary effects. The coarsest Cauchy grid has $51^3$ points and
covers the domain $x^i \in [-10M,10M]$. The coarsest characteristic
grid has $35^2 \times 31$ points. The extraction world-tube is at
$r=7M$.  The simulation is evolved until $t=100M$.

The CCE news converges to zero as it should. However, 
Zerilli extraction does not give a signal that converges to zero 
as grid resolution is refined. Instead, an erroneous non-trivial 
signal is computed. 

\begin{figure}[htbp]
  \centering
  \includegraphics[width=7.5cm]{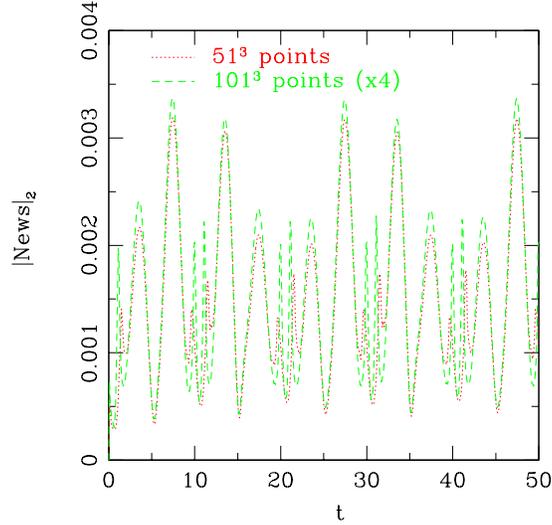}
  \caption{The scaled $L_2$ norm of the CCE news for a static spherically
    symmetric black hole in an oscillating frame. The CCE news 
    converges to zero at second order as it should.}
  \label{fig:News_bbh}
\end{figure}

\begin{figure}[htbp]
  \centering
  \includegraphics[width=7.5cm]{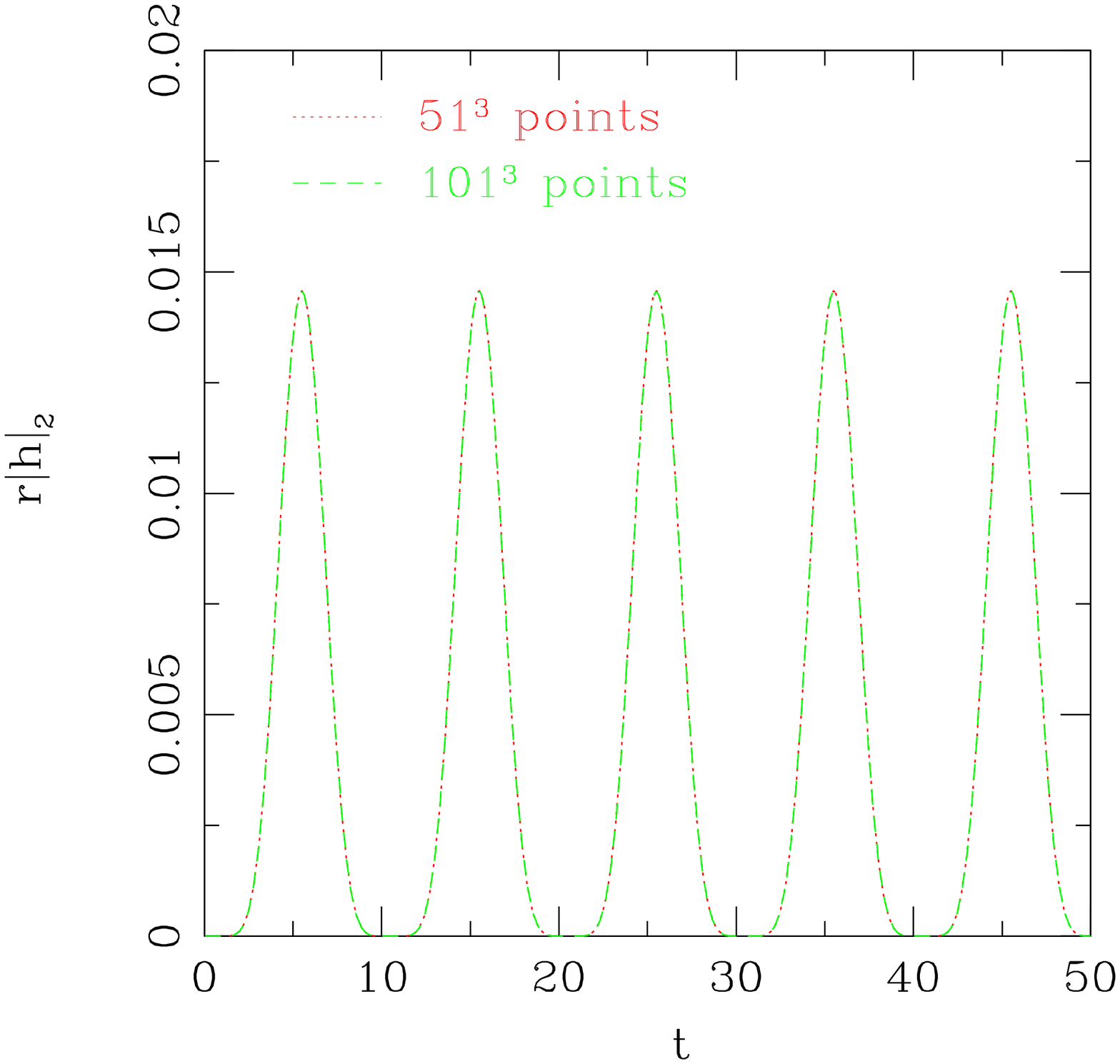}
  \includegraphics[width=7.5cm]{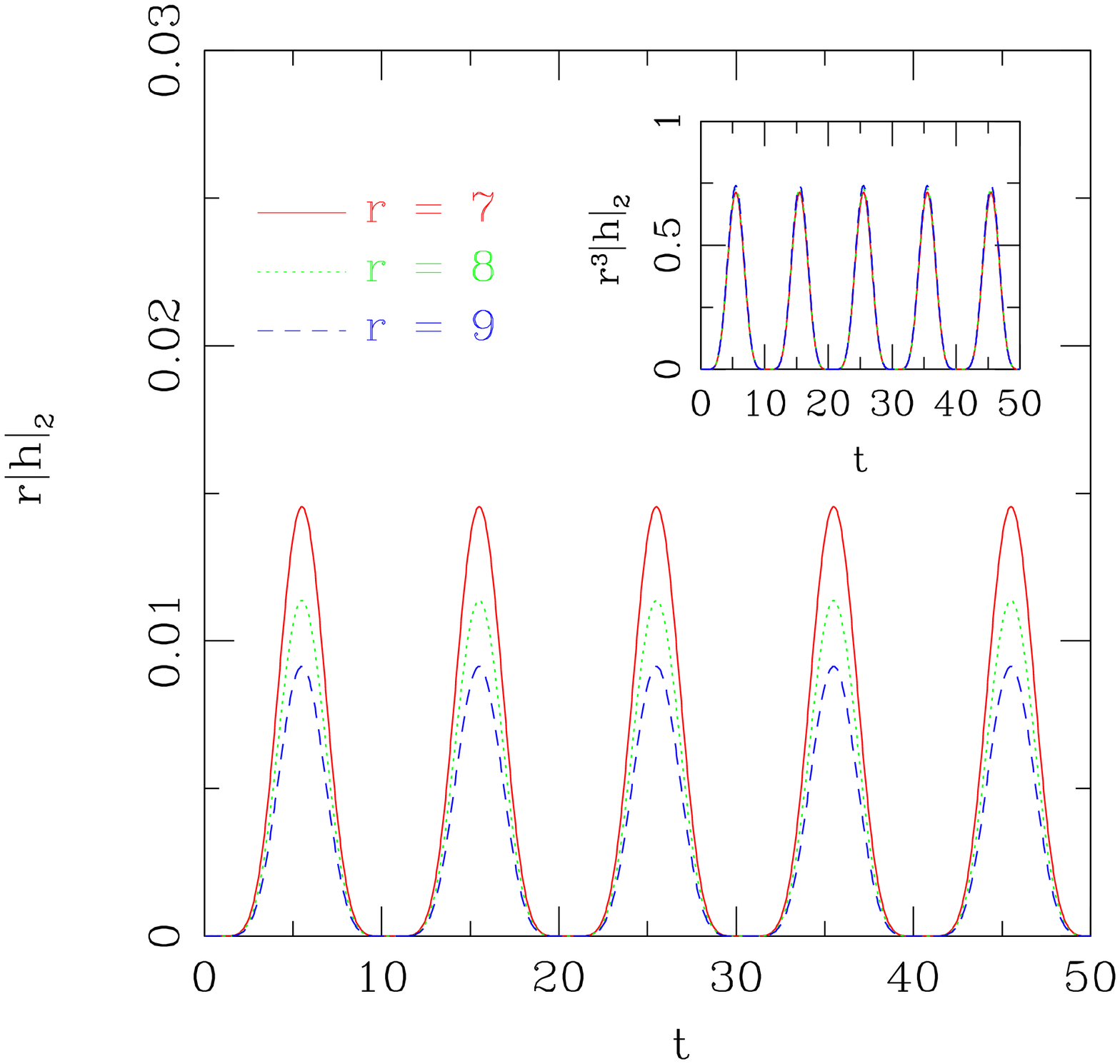}
  \caption{The wavesignal computed using Zerilli extraction for the
    static spherically symmetric black hole in an oscillating frame.
    The ``norm'' is computed as described in the text with the radial
    dependence removed. The left panel shows that as the resolution of
    both the Cartesian grid and extraction sphere are increased the
    signal does not converge to zero. The right panel shows how this
    erroneous non-trivial wavesignal decays as ${\cal O}(r^{-2})$ as 
    the extraction radius is increased. In the main plot the standard scaling
    of the signal is used. In the inset the curves are scaled by the
    extraction radius so as to overlay each other for the expected decay
    rate.}
  \label{fig:Zerilli_bbh}
\end{figure}

\subsection{Teukolsky Wave Test}
\label{sec:tests_teuk}
\subsubsection{Theoretical part}

 The Teukolsky solution~\cite{Nakao01, Kidder01a, Fiske05} to the linearized 
Einstein equation is a weak gravitational wave propagating through space, and 
represents one of the most valuable standard test cases for numerical codes 
that implement wave extraction techniques.

 The general form of the spacetime metric is
\begin{eqnarray} 
ds^2 & = &-dt^2 +(1+Af_{rr})dr^2 + 2Bf_{r\theta}rdrd\theta \nonumber \\
     & + & 2Bf_{r\phi}r \sin\theta dr d\phi + (1+Cf_{\theta \theta}^{(1)}
     + Af_{\theta \theta}^{(2)})r^2 d \theta ^2 \nonumber \\
     & + & 2(A-2C)f_{\theta  \phi} r^2 sin \theta d \theta d \phi 
     + (1+Cf_{\phi \phi}^{(1)} + Af_{\phi \phi}^{(2)})r^2 \sin^2 \theta d \phi^2
\end{eqnarray}
where the angular functions $f_{ij}$, corresponding to an $l=2,~m=0,
~spin-weight=2$ spherical harmonic, are
\begin{eqnarray}
& &f_{rr} = 2 - 3 \sin^2 \theta,\space 
f_{r\theta} = - 3 \sin \theta cos \theta, \space 
f_{r\phi} = 0 , \nonumber \\
& &f_{\theta \theta}^{(1)}= 3 sin^2\theta, \space
f_{\theta \theta}^{(2)} =  -1 , \nonumber \\
& &f_{\theta \phi} = 0 ,\space
f_{\phi \phi}^{(1)} = -f_{\theta \theta}^{(1)} , \space 
f_{\phi \phi}^{(2)} = 3 \sin^2 \theta -1 .
\end{eqnarray} 
 The functions A, B and C are given in terms of a free generating function 
 $ F( x ) = {\cal A} x e^{ - x ^ 2 / \lambda^2} / \lambda^2 $, where ${\cal A}$
 is the amplitude and $\lambda$ determines the width of the wave, by
\begin{eqnarray}
A & = & 3\big ( \frac{ d_x^2 F}{r^3} + \frac{3 d_x F}{r^4} 
      + \frac{3F}{r^5} \big ) 
\nonumber \\
B & = & -\big ( \frac{ d_x^3 F}{r^2} + \frac{3 d_x^2 F}{r^3} 
      + \frac{6 d_x F}{r^4} + \frac{6F}{r^5} \big ) 
\nonumber \\
C & =& \frac{1}{4}\big ( \frac{d_x^4 F}{r} + \frac{2 d_x^3 F}{r^2}+
      \frac{9 d_x^2 F}{r^3} + \frac{21 d_x F}{r^4} + \frac{21F}{r^5} \big ),
\end{eqnarray}
where $ d_x^n F$ denotes 
\begin{equation}
d_x^n F= \frac{d^nF(x)}{dx^n}.  
\end{equation}
 Here, $x=t-r$ corresponds to an outgoing wave. To obtain the ingoing 
wave solution coresponding to $x=t+r$, we change the sign in front of all the
terms with odd numbers of $F$.

 We consider a superposition of ingoing $t+r$ and outgoing $t-r$ waves
~\cite{Szilagyi00}, centered at the origin of the coordinate system at $t=0$, 
which provides a moment of time symmetry. This solution, after is linearized 
in amplitude, is implemented in the harmonic Abigel code. 

 We further give the analytical form of the Teukolsky wave signal.
The real part of the Bondi news function is proportional to the time 
derivative of the "plus" polarization mode of the gravitational wave signal.
The time derivative of the CCE news satisfies
\begin{equation}
\label{eq:NewsPsi}
\dot N = \lim_{r \rightarrow \infty} r \Psi_4 
\end{equation}
where $\Psi_4$ is the Newman-Penrose component of the Weyl tensor 
\begin{equation}
 \Psi_4 = - C_{\mu \rho \nu \tau} n^\mu {\bar m}^\rho n^\nu {\bar m}^\tau  
\end{equation}
 with $n^\nu$ an ingoing null vector, and $ m^\rho$ a complex 
 unit vector, oriented in the angular directions. 

 The connection between $m^\rho$ and the dyad vector $q^A$ on the unit sphere is
\begin{equation}
 m^A = \frac{1}{r} q^A .
\end{equation}
Moreover, $n^\mu \partial_\mu = \partial_x$.
A straightforward computation starting from Eq.~\eqref{eq:NewsPsi}, gives
\begin{equation}
 \dot N = - \frac{1}{ r} {\bar q}^A {\bar q}^B \partial_x^2 g_{A B}.
\end{equation}
 At infinity, in the linearized regime, only the ${\cal O}(1/r)$ terms in the 
Teukolsky functions $A$, $B$ and $C$ contribute to the news. 
The term that determines the gravitational wave signal is $d_x^4 F$.
We find
\begin{equation}
\label{eq:DotNews}
\dot N = -\frac{3 \sin^2 \theta }{4} \partial_t ^6 F.
\end{equation}
Alternatingly, computation of the linearized expression for the Bondi news 
from Eq.~\eqref{eq:News}, gives
\begin{equation}
\label{eq:NewsTeuk}
 N = -\frac{3 \sin^2 \theta }{4} \partial_t ^5 F.
\end{equation}

\subsection{Numerical part}
\label{sec:tests_teuk_num}

We give the metric specified by the Teukolsky solution analytically on
the Cauchy grid. Then data is extracted at the world-tube and evolved
by the CCE code.  We carry out a series of simulations,
varying both the location of the characteristic world-tube and the
radius of the Zerilli sphere. We use $80^3$, $120^3$, and $160^3$ 
gridpoints for the Cartesian Cauchy grid and $60^2\times 80$, $90^2\times 120$,
and $120^2 \times 160$ gridpoints for the null grid. The domain extends between 
$x^i \in [-15,15]$ and the simulations are run until $t=30$.

We study the dependence of the signal with the amplitude and conclude
that CCE code resolves correctly amplitudes of ${\cal A} \geq 10^{-8}$.
At smaller amplitudes, clean convergence behavior is contaminated by round-off 
error. We show results only for ${\cal A}=10^{-5}$, $\lambda = 1$.
We study also the dependence of the wave signal with the world-tube radius 
and conclude that the accuracy of the computed CCE news is preserved even for 
radii as small as $r=5$.
 
\begin{figure}[htbp]
  \centering
  \includegraphics[width=7.5cm]{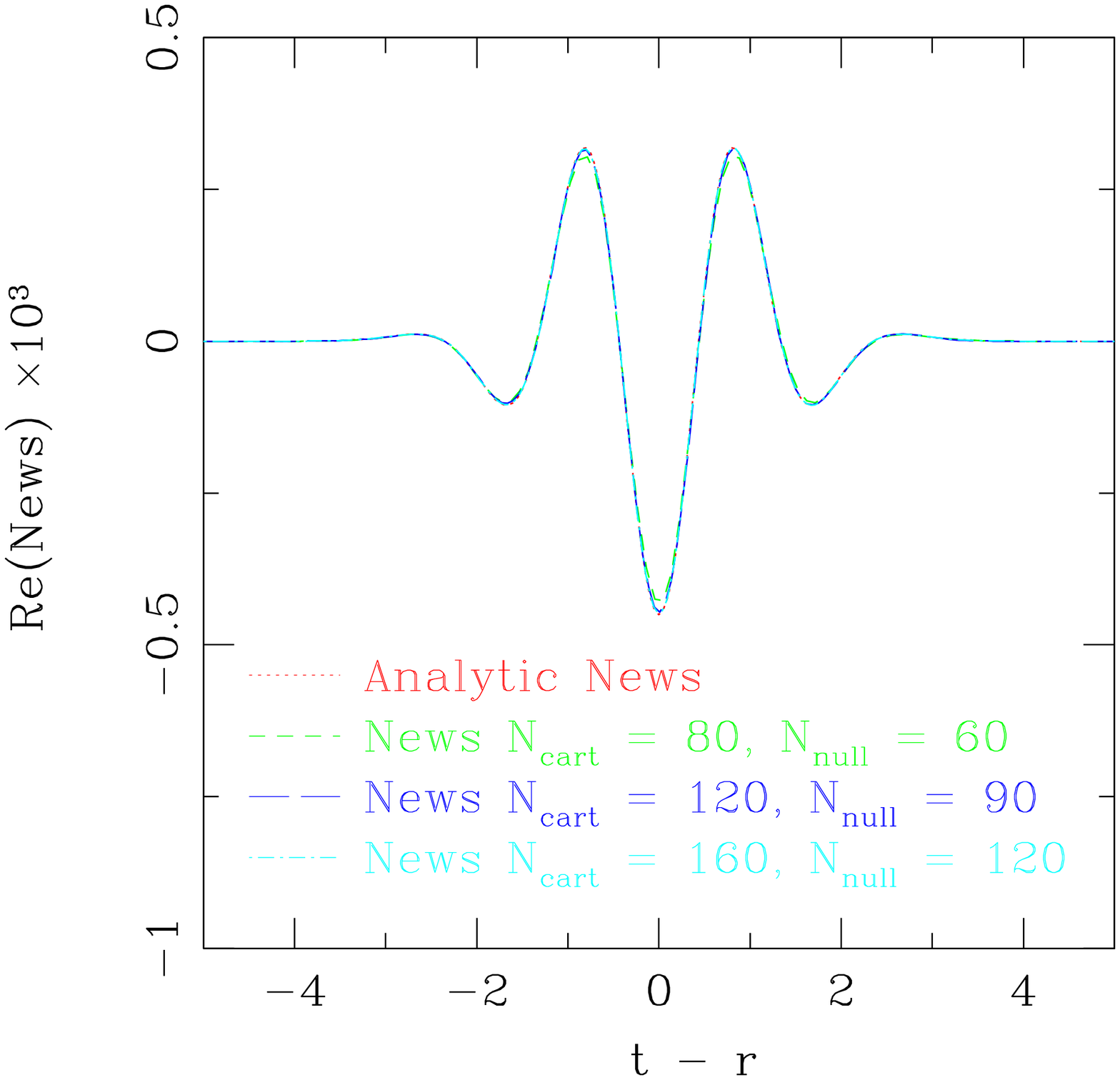}
  \includegraphics[width=7.5cm]{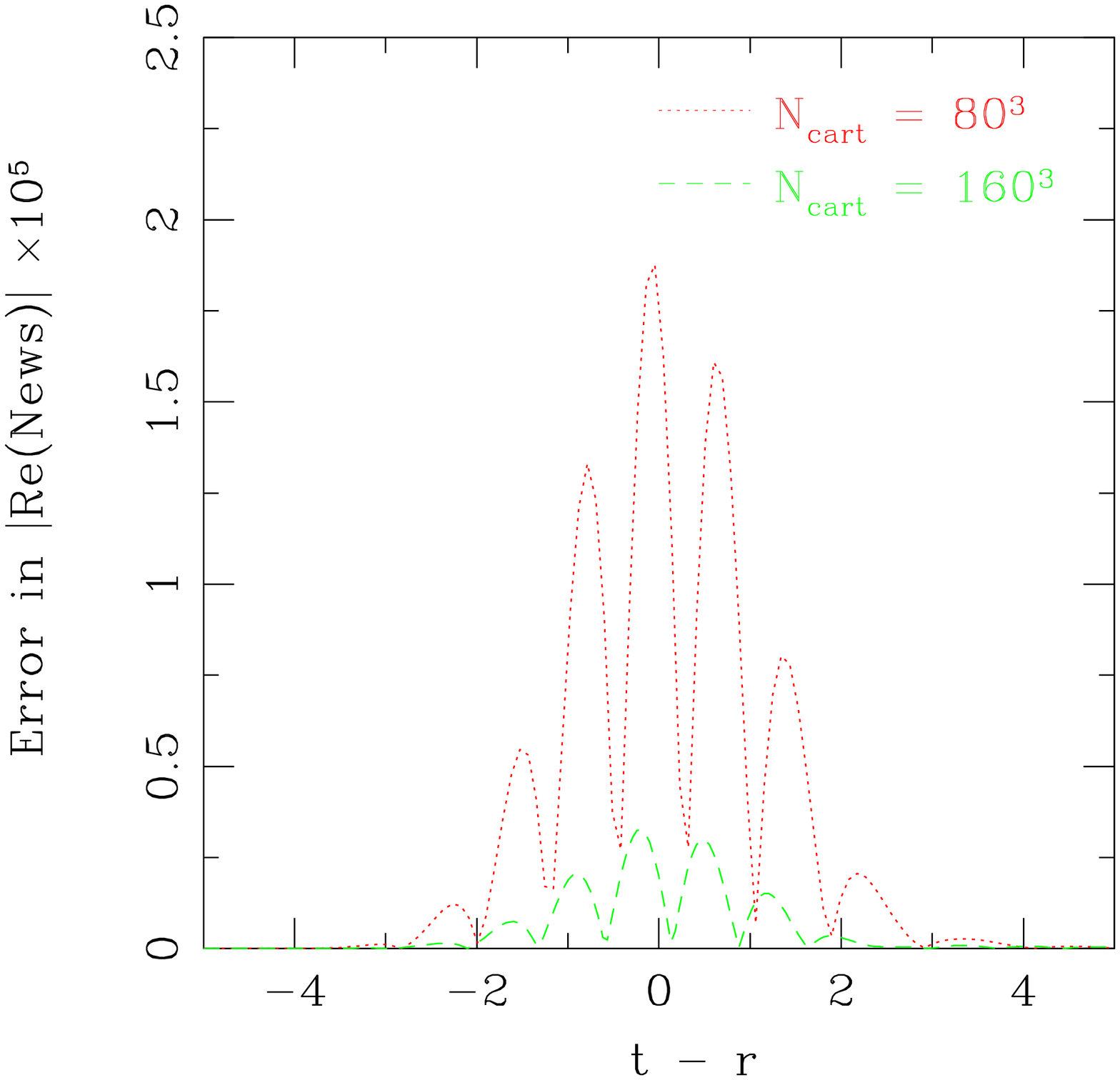}
  \caption{Left panel: the convergence of CCE news to the
    analytical solution. In this case the grid resolution was 
    (Cartesian / characteristic) $80^3 / 60^2 \times
    80$, $120^3 / 90^2 \times 120$ and $160^3 / 120^2 \times 160$
    respectively.
    Right panel: the absolute value of the error in CCE news for the grid 
    resolutions of $80^3 / 60^2 \times80$ and $160^3 / 120^2 \times 160$.
    }
  \label{fig:CCE_Conv}
\end{figure}

Fig.~\ref{fig:CCE_Conv} shows the convergence of the 
CCE news to the analytical solution, for a world-tube radius of $r=10$.
The convergence rate ${\tilde c_r}$ of the CCE news to the
analytical data, is given by
\begin{equation}
  {\tilde c_r} = \log_2 \big(
  \frac{||N_{80}-N_{ana}||}{||N_{160}-N_{ana}||}
  \big ),
\end{equation}
The convergence rate of the computed CCE news to the analytic value Eq.~\eqref{eq:NewsTeuk} at $t-r=0$, corresponding to the peak of the radiated 
signal, is 
\begin{equation}
{\tilde c_r} = 2.159.
\end{equation}
Hence, the CCE wavesignal is second order convergent and independent of the 
world-tube radius, as expected.

\begin{figure}[htbp]
  \centering
  \includegraphics[width=7.5cm]{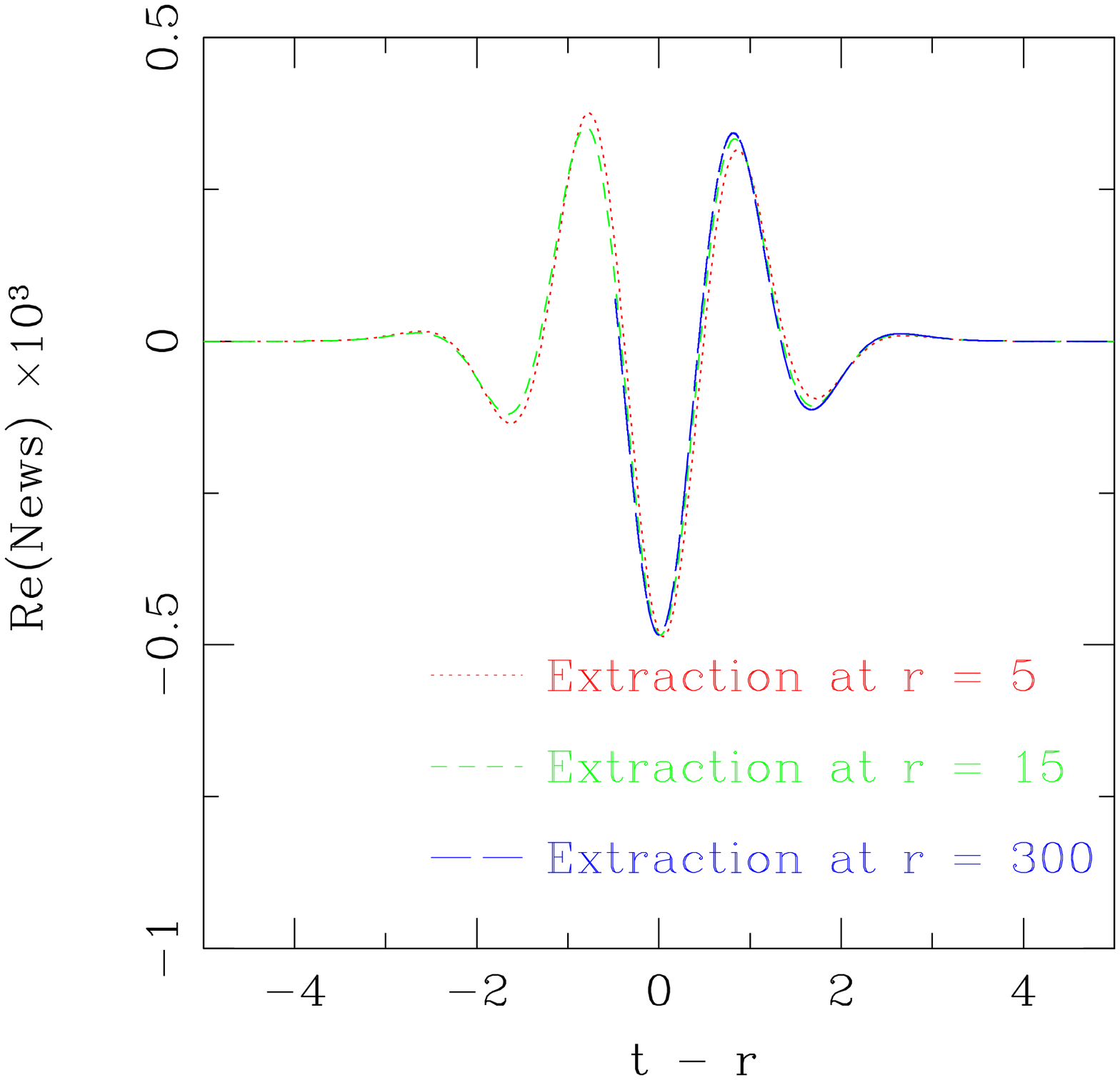}
  \includegraphics[width=7.5cm]{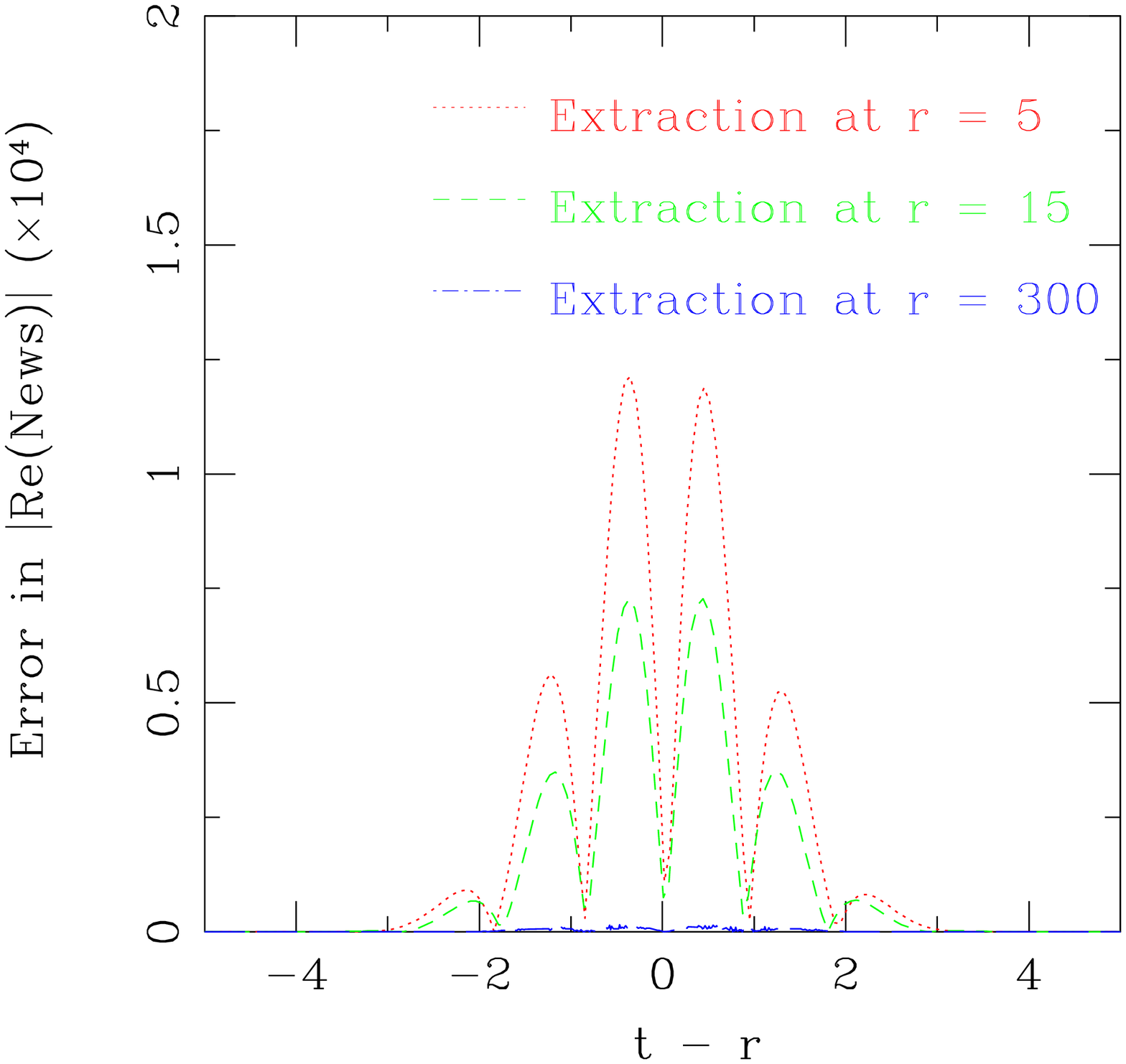}
  \caption{Left panel: the Zerilli news shows a dependence on the extraction 
           radius. The asymmetry decreases as the extraction radius increases.
           Right panel: the absolute value of the error in Zerilli
           news with radius (grid resolution of $120^3 / 90^2 \times 120$.}
  \label{fig:Zerilli_Radius}
\end{figure}

\begin{figure}[htbp]
  \centering
  \includegraphics[width=7.5cm]{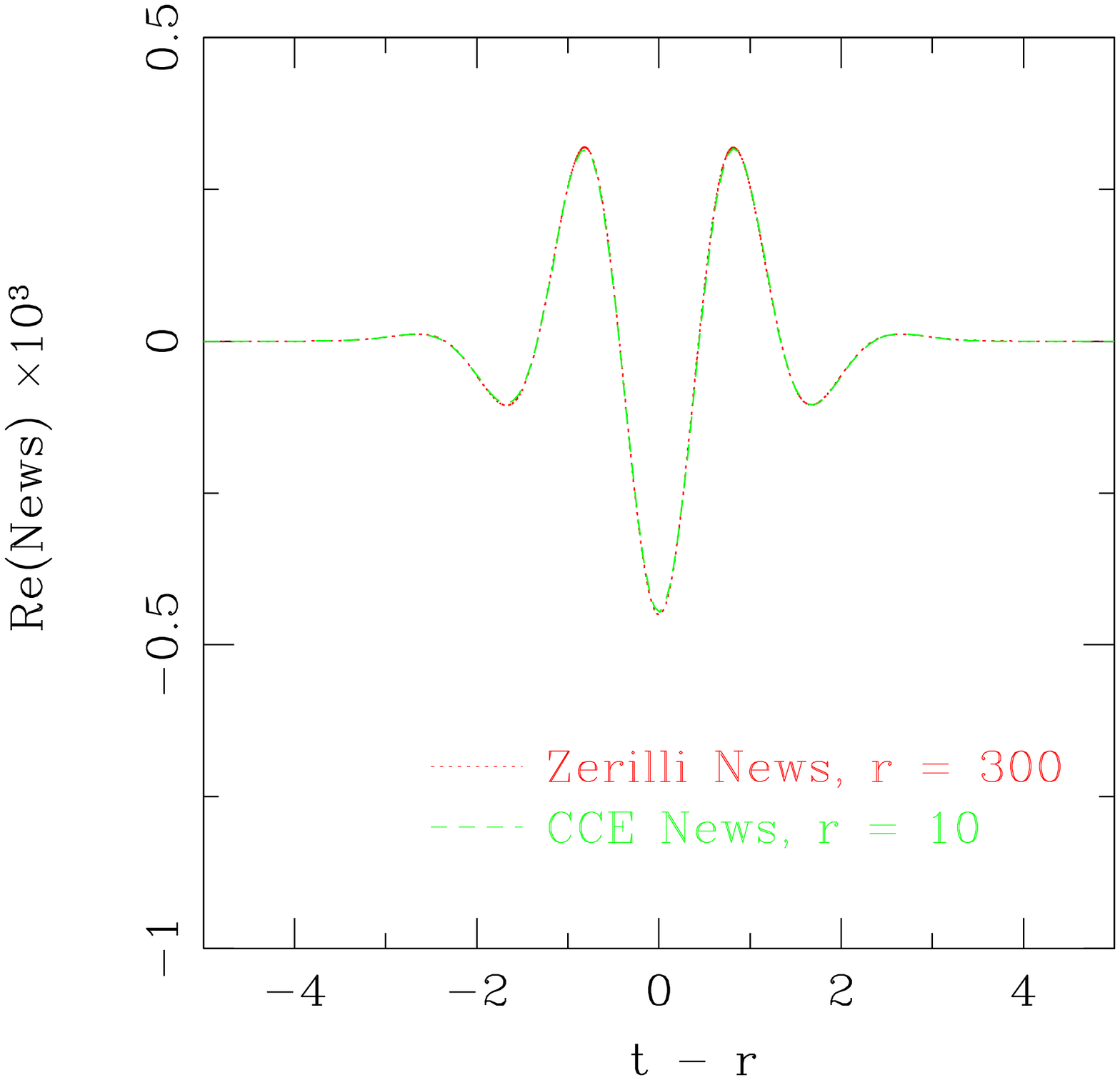} 
  \caption{ When the extraction radius is sufficiently large, 
            the comparison between the Zerilli and the CCE 
            news is in very good agreement.
}
  \label{fig:Zerilli_CCE}
\end{figure}

Fig.~\ref{fig:Zerilli_Radius} shows that for small extraction radii, the 
Zerilli waveform has a slight asymmetry, which is caused by the dependence 
of the Zerilli formalism upon the extraction radius. We have to increase the 
extraction radius in order to decrease this error. 
Because the error does not fall off sufficiently fast with radius in the 
near zone, where we can realistically carry out the simulation, we instead 
analytically compute the Zerilli news at the extraction radius $r=300$.
Fig.~\ref{fig:Zerilli_CCE} demonstrates that the agreement between the 
computed CCE news at small radii and the analytical Zerilli news at big radii, 
is very good. 

\section{Conclusions}
\label{sec:sec5}

We have demonstrated the accuracy of Cauchy-characteristic extraction
in 3D numerical relativity. The interface at the extraction world-tube
does not introduce significant error with either the BSSN~\cite{Alcubierre02a}
or harmonic~\cite{Szilagyi02b,Szilagyi02a} formulations.

The CCE news is stable against small perturbations on the Cauchy grid, 
as shown in Sec.~\ref{sec:tests_rand}, and the level of truncation error 
caused by transforming from the Cauchy to the characteristic variables 
is small, as shown in Sec.~\ref{sec:tests_flat}. 
In a non-trivial black hole spacetime, CCE performs as expected. Second 
order convergence is found in the moving Schwarzhild black hole test in 
Sec.~\ref{sec:tests_bh_bbh} and in the early time behavior of the BSSN test 
in Sec.~\ref{sec:tests_bh_fix}.
However, as seen in the late time behavior in Sec.~\ref{sec:tests_bh_fix}, 
the effect of improper BSSN outer boundary conditions on the Cauchy grid 
is clearly visible in the extracted CCE news.

In the linearized Teukolsky wave test in Sec.~\ref{sec:tests_teuk}, we 
demonstrate that the computed CCE news converges to the analytic Teukolsky 
waveform and does not depend on the world-tube radius. The accuracy of the 
CCE news is preserved even for small radii, while the Zerilli news is 
affected by near zone error. When the extraction radius is sufficiently large, 
both Zerilli and CCE give excellent results. 
The advantage of CCE over the Zerilli method is that the extracted 
CCE news does not depend on the extraction radius. 

Cauchy-characteristic extraction can be seen either as a first step
towards a full Cauchy-characteristic matching code or as an improved
gravitational wave extraction method in its own right. The results of
this paper show that as a stand alone wave extraction method, it
produces the correct results in situations where other methods, such
as Zerilli extraction, may fail. Such a situation is seen in
Sec.~\ref{sec:tests_bh_bbh}, where the Zerilli method fails to converge as 
a result of the pure gauge motion of the Schwarzchild metric. 

Finally, these tests also show the need for improvement of the current
implementation of the code. Any angular dependence in the solution
whether through genuine physics, gauge effects, or due to the
imposition of boundary conditions on the cubical Cauchy boundary,
leads to short wavelenght error that is poorly resolved by the CCE code. 
This is particularly noticeable for features in the region where the 
stereographic patches overlap. Improvements in this area of the implementation 
will enhance the accuracy of the code in extracting the gravitational news 
from astrophysical simulations requiring high resolution. To this end, we 
are investigating the use of different multiple patch implementations such
as~\cite{Thornburg2004:multipatch-BH-excision}.

\begin{acknowledgments}

We are grateful to Nigel Bishop for help at the start of this project.
Also, we are thankful to Jeff Winicour for his support throughout
the project as well for his careful reading of the manuscript.
Computer simulations  were done at the PEYOTE cluster of the
Albert Einstein Institut in Golm, Germany
and at the Pittsburgh Supercomputing Center under grant PHY040015P.
While importing the extraction code into the Cactus infrastructure
we have benefited from the support of the Cactus teams at AEI and LSU, 
for which we are grateful.

We thank LSU for its hospitality. MB was supported by the National Science 
Foundation under grant PHY-0244673 to the University of Pittsburgh.
BSz was partially supported by the National Science Foundation under
grant PHY-0244673 to the University of Pittsburgh.
IH was partially supported by PPARC grant PPA/G/S/2002/00531.
YZ was partially supported by  the NASA Center for Gravitational
Wave Astronomy at The University of Texas at Brownsville (NAG5-13396) and by
NSF grants PHY-0140326 and PHY-0354867.

\end{acknowledgments}

\bibliographystyle{apsrev}
\bibliography{bibtex/references}

\end{document}